\documentstyle[12pt]{article}
\def\b{\tilde{b}}
\def\a{\tilde{a}}
\def\S{\tilde{S}}
\def\A{{\cal A}}
\def\B{{\cal B}}
\def\C{{\cal C}}
\def\D{{\cal D}}
\def\d{\hat{{\cal D}}}
\def\P{{\cal P}}
\def\1{{\bf 1}}

\begin{document}
\begin{titlepage}
\title{Integrable open-boundary conditions\\
 for the supersymmetric
t-J model.\\
The quantum group invariant case.}
\author{A. Gonz\'alez--Ruiz \thanks{Work supported by the Spanish
M.E.C. under grant AP90 02620085}\\
{\it  Departamento
de F\'{\i}sica Te\'orica} \\
{\it Universidad Complutense}\\
{\it 28040 Madrid,
ESPA\~NA}}
\date{}
\maketitle

\begin{abstract}
We consider integrable open--boundary conditions for the supersymmetric t--J
model commuting with the number operator $n$ and $S^{z}$. Four families, each
one  depending on two arbitrary parameters, are found.  We find the relation
between Sklyanin's method of constructing open boundary conditions and the
one for the quantum group invariant case based on Markov traces.  The
eigenvalue problem is solved for the new cases by generalizing the Nested
Algebraic Bethe ansatz of the quantum group invariant case (which is
obtained as a special limit). For the quantum group invariant case the Bethe
ansatz states are shown to be highest weights of $spl_{q}(2,1)$.
\end{abstract}

\vskip-18.0cm
\rightline{{\bf hep-th 9401118}}
\rightline{{\bf F.T/U.C.M-94/1}}
\rightline{{\bf January 1994}}
\vskip3cm

\end{titlepage}

\begin{section}{Introduction}
During the last years there has been an increasing activity on the study of
systems in the finite interval. For the case of one dimensional models
 integrable by the quantum inverse scattering method the pioneering works of
 Cherednik and Sklyanin \cite{ch,sk} were the starting point. There, the so
called reflection equations, appeared as the new ingredient to the
Yang-Baxter equation when dealing with independent boundary conditions. A
lot of integrable models arised using this method, see for example
\cite{mn,ks}. Some of them have been shown to have physical
applications as for example the Hofstadter problem \cite{wz} or the
reaction diffusion equations \cite{adhr}. Some of these models have been
solved and other would need non trivial generalizations of the
methods developed for periodic and twisted boundary conditions  (for
a review see \cite{hr}). Certain limits of this open boundary conditions
where shown to give quantum group invariant transfer matrices and
hamiltonians \cite{ps,qba1,mn,ks,ro}. \\
On the other hand the t-J model has attracted much interest in connection
with high-$T_{c}$ superconductivity. This model is obtained from the
Hubbard model as an effective hamiltonian for the low-energy states in the
strong correlation limit. In this limit double occupancy of fermions is
forbidden, leading to only three possible states at each lattice site. For
a given ratio of the coupling constants the model becomes supersymmetric. In
one dimension the model is exactly solvable by means of the nested Bethe
ansatz method, see for example \cite{ttjj}. Recently the quantum inverse
scattering method was used to solve this model \cite{kore} and the
completeness of the Bethe states was shown \cite{fk2}. Also a $spl_{q}(2,1)$
invariant t-J model
 was proposed and solved by a generalization
of the nested Bethe ansatz method to quantum group invariant open boundary
conditions \cite{fk1}. This method has been also generalized for the case of
the $SU_{q}(n)$
 invariant chains \cite{suq}. Up to this moment for the susy t-J model only
the periodic and the quantum group invariant cases have been solved by means
of the nested Bethe ansatz.\\
Some fundamental questions remain open, among them: the
solution of the generic open chain with independent boundary conditions,
the highest weight property for the quantum group invariant chain, the
problem of completeness of Bethe states and the relation of the Markov trace
method for finfing quantum group invariant chains with the method of
reflection equations.\\
It is interesting to find the widest possible
class of open boundary conditions compatible with the integrability of the
model. This terms could give count of impurities or magnetic fields located
at the boundaries. In the present work we find new integrable boundary
conditions for the supersymmetric t-J hamiltonian. We look only for
solutions commuting with the operators $n$ and ${\bf S^{z}}$ because these
are the ones keeping the structure of the low-energy states. Other boundary
conditions would give rise to low-energy states of the domain wall type
\cite{smr,ks}. These new conditions survive in the limit of cero anisotropy,
which is not the case for those of the quantum group invariant case. We find
the neccesary generalization of the  nested Bethe ansatz construction to the
case of independent boundary conditions in the edges of the chain.\\
For the
quantum group invariant chain we prove the highest weight property for the
Bethe eigenvectors. This is the first proof of this property in an invariant
chain  with respect to a quantum group of rank greater than one and is
the first step in order to treat the completness problem (recently this
property has been also shown to hold in the $SU_{q}(n)$ invariant case
\cite{hwp}). We will also see that there is a degeneration in the
eigenvalues.\\
A general method based on Markov traces  to find quantum
group invariant chains has ben proposed recently \cite{fk1,kk}. We find the
connection with the method of reflection equations modified for non
symmetric $S$-matrices in \cite{mn}. The proof of the commutativity of the
transfer matrices and of the quantum group invariance is immediate from this
point of view.\\      The paper is organized as follows. In section 2 all the
diagonal solutions to the reflection equations are found, giving the
relation between the transfer matrix and the new integrable families of
hamiltonians. We also discuss the relation of Markov traces with
reflection equations in the quantum group invariant case. In section 3 we
diagonalize the transfer matrices using the quantum inverse scattering method
generalizing Sklyanin's approach to open boundary conditions for the Nested
Bethe Ansatz construction. Section 4 is devoted to probe the highest weight
property for the Bethe vectors in the quantum group invariant case. Section
5 contains a summary of the main results and some further possible
investigations.

\end{section}

\begin{section}{Integrable open--boundary
conditions}

As is by now well known the supersymetric t--J model is related to the
graded 15 vertex model hamiltonian, see for example \cite{fk1,fk2}. This
is is a vertex model with three states per link which can be bosonic or
fermionic. The matrix of vertex weights has the form:

\begin{eqnarray}
S^{ab}_{cd}(v)=\left(\begin{array}{ccccccccc}
a&0&0&0&0&0&0&0&0\\
0&b&0&c_{-}&0&0&0&0&0\\
0&0&b&0&0&0&c_{-}&0&0\\
0&c_{+}&0&b&0&0&0&0&0\\
0&0&0&0&a&0&0&0&0\\
0&0&0&0&0&b&0&c_{-}&0\\
0&0&c_{+}&0&0&0&b&0&0\\
0&0&0&0&0&c_{+}&0&b&0\\
0&0&0&0&0&0&0&0&w
\end{array}\right), \label{s}
\end{eqnarray}

where $a$, $b$, $c$ and $d$ are indexes running from 1
to 3 and\\
$a=\sin(v+\gamma)$, $b=\sin v$, $c_{+}=e^{iv}\sin\gamma$,
$c_{-}=e^{-iv}\sin\gamma$, $w=\sin(-v+\gamma)$.\\
We have adopted the convention of making fermionic the third state.\\
This matrix is a trigonometric solution to the quantum Yang-Baxter equation:

\begin{eqnarray}
S_{12}(u)S_{13}(u+v)S_{23}(v)=S_{23}(v)S_{13}(u+v)S_{12}(u),\nonumber
\end{eqnarray}
defined in the space $V_{1}\otimes V_{2}\otimes V_{3}$ with the standard
notation $S_{ij}\in End(V_{i}\otimes V_{j})$.\\
This $S$-matrix does not enjoy P and T symmetry but just PT invariance:

\begin{eqnarray}
PS_{12}P:=S_{21}=S_{12}^{t_{1}t_{2}}\nonumber  .
\end{eqnarray}

It is not crossing invariant either but it obeys the weaker property
\cite{n}:

\begin{eqnarray}
\left[\left\{\left[S_{12}(v)^{t_{2}}\right]^{-1}\right\}^{t_{2}}
\right]^{-1}
=   L(v,\gamma) M_{2} S_{12}(v +2\eta) M_{2}^{-1},
\label{sm}
\end{eqnarray}
where $L(v,\gamma)$ is a c-number function, $\eta$ a
constant and $M$ a
symmetry of the $S$ -matrix:

\begin{eqnarray}
[M_{1} \otimes M_{2}, S_{12}(v)] = 0.
\nonumber
\end{eqnarray}

We find by direct calculation from (\ref{s}) and (\ref{sm}):

\begin{eqnarray}
\eta & = & \frac{\gamma}{2}\nonumber\\
M & = & \left(\begin{array}{ccc}
1&0&0\\
0&q^{2}&0\\
0&0&-q^{2}
\end{array}\right);\;\;q=e^{i\gamma},
\label{m}\\
L(v,\gamma) & = & -\frac{b(v)}{w(v)}. \nonumber
\end{eqnarray}

When the weak condition (\ref{sm}) holds, if looking for integrable boundary
conditions, it is neccesary to solve for $K^{\pm}(v)$ the following
 reflection equations \cite{mn}:

\begin{eqnarray}
S_{12}(u-v)K_{1}^{-}(u)S_{21}(u+v)K_{2}^{-}(v)=\nonumber\\
K_{2}^{-}(v)S_{12}(u+v)K_{1}^{-}(u)S_{21}(u-v), \label{sk1}
\end{eqnarray}

\begin{eqnarray}
S_{12}(-u+v)K^{+}_{1}(u)^{t_{1}}
M_{1}^{-1} S_{21}(-u-v-2\eta)M_{1}
K^{+}_{2}(v)^{t_{2}}=\nonumber\\
 K^{+}_{2}(v)^{t_{2}} M_{1} S_{12}(-u-v-2\eta)
M_{1}^{-1} K_{1}^{+}(u)^{t_{1}}  S_{21}(-u+v).\label{sk2}
\end{eqnarray}

There is an automorphism between $K^{-}$ and $K^{+}$:

\begin{equation}
K^{+}(v) = K^{-}(-v-\eta)^{t} M.\label{aut}
\end{equation}

As explained in sect. 1, it is interesting to find open boundary conditions
commuting with the operators $n$ and $S^{z}$ which are diagonal. For that
reason  we will only look for diagonal solutions to the equation (\ref{sk1})
:

\begin{equation}
K^{-}(v)_{ab}=K^{-}_{a}(v)\delta_{ab}\nonumber
\end{equation}

Inserting this in equation (\ref{sk1}) the only nontrivial
equations are:

\begin{eqnarray}
&&\sin(u+v) [K_{b}^{-}(u)
K_{a}^{-}(v) e^{sign(a-b)(u-v)}  -
K_{a}^{-}(u) K_{b}^{-}(v)
e^{-sign(a-b)(u-v ) }]\nonumber\\
&&+\sin(u-v) [K_{a}^{-}(u)
K_{a}^{-}(v)e^{-sign(a-b)(u+v)} -
K_{b}^{-}(u) K_{b}^{-}(v)
e^{sign(a-b)(u+v) }]  =  0,\nonumber\\
&&a,b=1,2,3. \nonumber
\end{eqnarray}

Deriving this equation with respect to $u$ and making $u=0$ the solutions
to these equations are found to be:

\begin{eqnarray}
&&K_{a}(v)=e^{iv}\sin(\xi_{-}-v)\nonumber\\
&&K_{b}(v)=e^{-iv}\sin(\xi_{-}+v);\;\;\;a>b\nonumber
\end{eqnarray}
with $\xi_{-}$ an arbitrary parameter.\\
This gives two families of solutions for the equation (\ref{sk1}) and using
the automorphism (\ref{aut}) the same number for the equation
(\ref{sk2}), these are:

\begin{eqnarray}
K_{a}^{-}(v)&=&\frac{1}{\sin\xi_{-}}\left(\begin{array}{ccc}
e^{-iv}\sin(\xi_{-}+v)&0&0\\
0&e^{-iv}\sin(\xi_{-}+v)&0\\
0&0&e^{iv}\sin(\xi_{-}-v)
\end{array}\right),\label{kma}\\
K_{b}^{-}(v)&=&\frac{1}{\sin\xi_{-}}\left(\begin{array}{ccc}
e^{-iv}\sin(\xi_{-}+v)&0&0\\
0&e^{iv}\sin(\xi_{-}-v)&0\\
0&0&e^{iv}\sin(\xi_{-}-v)
\end{array}\right),\label{kmb}\\
K_{a}^{+}(v)&=&\frac{1}{\sin\xi_{+}}\left(\begin{array}{ccc}
e^{iv}\sin(\xi_{+}-v)&0&0\\
0&q^{2}e^{iv}\sin(\xi_{+}-v)&0\\
0&0&-q e^{-iv}\sin(\xi_{+}+v+\gamma)
\end{array}\right),\nonumber\\
K_{b}^{+}(v)&=&\frac{1}{\sin\xi_{+}}
\left(\begin{array}{ccc}
e^{iv}\sin(\xi_{+}-v)&0&0\\
0&q e^{-iv}\sin(\xi_{+}+v+\gamma)&0\\
0&0&-q e^{-iv}\sin(\xi_{+}+v+\gamma)
\end{array}\right),\nonumber
\end{eqnarray}
where $\xi_{+}$ and $\xi_{-}$ are arbitrary independent parameters. As
 it is obvious from equations (\ref{sm},\ref{sk1},\ref{sk2}) these solutions
can be multiplied by arbitrary factors, these have been chosen in
(\ref{kma}),(\ref{kmb}) in order to have $K^{-}(0)=1$. It is important to
note at this point that the fact of having several families of diagonal
solutions is a caracteristic of $S$-matrices corresponding to algebras with
rank bigger than one \cite{ks}. The number of independent solutions equals
the rank of the algebra as in the $A_{n-1}$ \cite{ks}.\\
For fixed boundary
conditions described by the matrices $K^{\pm}(v)$, one uses the monodromy
matrix:

\begin{eqnarray}
U_{ab}(v)=\sum_{cd}T_{ac}(v)K^{-}_{cd}(v)T_{db}^{-1}(-v).\label{mon}
\end{eqnarray}

Where $T_{ac}(v)$ is the standard monodromy matrix for a $L\times L$ square
lattice defined as the matrix product over the $S$'s:

\begin{eqnarray}
T_{ab(c)}^{\;\;(d)}(v)=S^{ad_{1}}_{b_{2}c_{1}}(v)
S^{b_{2}d_{2}}_{b_{3}c_{2}}(v)
S^{b_{3}d_{3}}_{b_{4}c_{3}}(v)...S^{b_{L}d_{L}}_{bc_{L}}(v),\nonumber
\end{eqnarray}
indexes in parenthesis act in the quantum space ${\bf C^{3}}\times
{\bf C^{3}}\times ...{\bf C^{3}}$ and $a$ and $b$ in the horizontal
auxiliary  space $C^{3}$ in the usual convention.\\
The operator $T^{-1}(v)$ is the inverse of $T(v)$ in both, horizontal and
 the quantum space and is given by:

\begin{eqnarray}
T_{ab(c)}^{-1(d)}(v)=\S^{b_{2}d_{1}}_{bc_{1}}(v)
\S^{b_{3}d_{2}}_{b_{2}c_{2}}(v)
\S^{b_{4}d_{3}}_{b_{3}c_{3}}(v)...\S^{ad_{L}}_{b_{L}c_{L}}(v),\nonumber
\end{eqnarray}
where:

\begin{eqnarray}
\S^{ab}_{cd}(v)=\frac{S^{ba}_{dc}(-v)}{\sin(\gamma+v)\sin(\gamma-v)}.
\nonumber
\end{eqnarray}

The elements of $\S$ will be denoted with a tilde "$\;\tilde{ }\;$".\\
The operator $U$ can be represented as a $3\times 3$ matrix of operators
acting in the quantum space:

\begin{eqnarray}
U=\left(\begin{array}{ccc}
\A&\B_{2}&\B_{3}\\
\C_{2}&\D_{22}&\D_{23}\\
\C_{3}&\D_{32}&\D_{33}
\end{array}\right).\label{u}
\end{eqnarray}

This operator matrix  satisfies the reflection equation
\cite{sk}:

\begin{eqnarray}
S_{12}(u-v)U_{1}^{-}(u)S_{21}(u+v)U_{2}^{-}(v)=\nonumber\\
U_{2}^{-}(v)S_{12}(u+v)U_{1}^{-}(u)S_{21}(u-v) .\label{sku1}
\end{eqnarray}

The fixed boundary condition transfer matrix is then defined as:

\begin{eqnarray}
t(v)=\sum_{ab} K_{ab}^{+}(v)U_{ba}(v).\label{tr}
\end{eqnarray}

Thanks to equations (\ref{sm},\ref{sk2},\ref{sku1},\ref{tr}) the transfer
matrix $t(v)$ defines a one parameter family of commuting operators
\cite{mn,sk}:

\begin{equation}
[t(v),t(u)]=0
,\nonumber
\end{equation}
showing the integrability of the model.\\
The transfer matrix $t(v)$ is related to a quantum one
dimensional hamiltonian through its first derivative in the usual way
\cite{sk}:

\begin{equation}
\dot{t}(0)=-\frac{1}{4}\;\sin\gamma\; H\;\; tr K^{+}(0)+tr \dot{K}^{+}(0),
\label{th}
\end{equation}
where $H$ is the hamiltonian given by:

\begin{equation}
H=\sum^{L-1}_{j=1}h_{j,j+1}+\frac{1}{2} \dot{K}^{-}_{1}(0)+
\frac{tr[K^{+}_{0}(0)h_{L0}]}{tr[K^{+}(0)]}.
\label{H}
\end{equation}

The subscript $0$ means the horizontal or auxiliary space, and the two sites
hamiltonian $h$ is given by:

\begin{equation}
h_{j,j+1}=P\;\dot{S}_{j,j+1}(0).\nonumber
\end{equation}

We see from (\ref{H}) that the efect of the previous construction in the
hamiltonian is to add terms depending on the matrices $K^{\pm}$ on the
edges of the chain. As we have two families of solutions for the
reflection matrices at each boundary, and they are independent, we will
have four kinds of boundary terms.\\
Using (\ref{H}) and omiting  a term proportional to the identity operator we
obtain the following hamiltonians:

\begin{eqnarray}
H&=&-\P\left\{\sum_{j=1}^{L-1}\sum_{\sigma}\left(c_{j,\sigma}^{\dagger}
c_{j+1,\sigma}+c_{j,\sigma}c_{j+1,\sigma}^{\dagger}\right)\right\}\P
\nonumber\\
&-&2\sum_{j=1}^{L-1}\left(S_{j}^{x}S_{j+1}^{x}+S_{j}^{y}S_{j+1}^{y}+
\cos\gamma\;\left(S_{j}^{z}S_{j+1}^{z}-\frac{n_{j}n_{j+1}}{4}\right)\right)
-\cos\gamma\sum_{j=1}^{L}n_{j}\nonumber\\
&+&i\sin\gamma\;(n_{1}-n_{L})-
i\sin\gamma\sum_{j=1}^{L-1}(n_{j}S^{z}_{j+1}-S^{z}_{j}n_{j+1})\nonumber\\
&+&i\sin\gamma \;H^{b}_{\alpha\beta},\;\;\alpha,\beta=a,b\label{htj}\\
H^{b}_{aa}&=&(\cot\xi_{-}-1)n_{1}-\left(\cot(\xi_{+}-\gamma)-1\right)n_{L}
\nonumber\\
H^{b}_{ab}&=&(\cot\xi_{-}-1)(S^{z}_{1}-n^{h}_{1}/2)-
\left(\cot(\xi_{+}-\gamma)-1\right)n_{L}
\nonumber\\
H^{b}_{ba}&=&(\cot\xi_{-}-1)n_{1}-(\cot\xi_{+}-1)(S^{z}_{L}-n^{h}_{L}/2)
\nonumber\\
H^{b}_{bb}&=&(\cot\xi_{-}-1)(S^{z}_{1}-n^{h}_{1}/2)-
(\cot\xi_{+}-1)(S^{z}_{L}-n^{h}_{L}/2).\nonumber
\end{eqnarray}

This is the supersymmetric t-J hamiltonian with four different kinds
of integrable boundary conditions depending each one on two arbitrary
parameters $\xi_{\pm}$. The operators $c^{(\dagger)}_{j\pm}$ are spin up or
down annihilation (creation) operators. The ${\bf S}_{j}$ are spin matrices
and $n_{j}$ ($n^{h}_{j}$) the occupation number of electrons (holes) at
lattice site $j$. The operator
$\P=\prod_{j=1}^{L}(1-n_{j\uparrow}n_{j\downarrow})$ forbids the double
occupancy of electrons at one lattice site. There are only three
possibilities $(1,2,3)=(\uparrow,\downarrow,0)$, an electron with spin up,
down or a hole. The boundary hamiltonians $H^{b}_{\alpha\beta}$ correspond
to choosing the families $\alpha$ for the $K^{+}$ matrix and $\beta$ for the
$K^{-}$.\\
The previous boundary conditions can be interpreted as the effect of
impurities or magnetic fields located at the edges of the chain.\\
There
appear in this hamiltonian imaginary boundary terms as in the case of the
open XXZ hamiltonian \cite{sk,ks}. But in this case imaginary
terms  appear also in the bulk part, this was first noticed for the
$SU_{q}(n)$ invariant models in \cite{ks} and in the case of the quantum
group invariant t-J model in \cite{fk1}. This seems to be a characteristic
of models with underlying quantum group of rank larger than one. For this
case, the imaginary bulk part is relevant for configurations
$\uparrow\downarrow$ or $\downarrow\uparrow$ which are separated by holes.
 This term also breaks the spin parity of the model. At one electron per site
(half filling) the imaginary bulk term in (\ref{htj}) reduce to that of the
XXZ model.\\
Although the hamiltonians obtained are non hermitean they have real
eigenvalues in some cases. In the quantum group invariant cases the
eigenvalues are real as shown in \cite{fk1,suq}. For the general boundary
conditions discussed in this article this is still true for the cases
$(\alpha,\beta)=(a,a),(b,b)$ as we will see in next section. In this
cases the hamiltonians become hermitean under an auxiliary scalar product
as shown in \cite{fk1} for the quantum group invariant case. In the
hiperbolic regime the eigenvalues are real in all the cases. In the
rational limit (the $spl(2,1)$ t-J model with open boundary conditions) all
the cases give hermitian hamiltonians.\\
We will now make some comments
concerning the quantum group invariant case. The $spl_{q}(2,1)$ invariant
hamiltonian of reference \cite{fk1} is obtained after taking in (\ref{htj})
the limit $\Xi_{\pm}=e^{i\xi_{\pm}}\rightarrow\infty$. When looking in this
limit to the reflection matrices  we have for the wo families of solutions
$K^{-}=1$ and:

\begin{eqnarray}
 K^{+}=M=\left(\begin{array}{ccc}
1&0&0\\
0&q^{2}&0\\
0&0&-q^{2}
\end{array}\right).\nonumber
\end{eqnarray}

Then we recover the construction of Foerster and Karowski defining the
transfer matrix as the Markov trace associated with the superalgebra
$spl_{q}(2,1)$ of the monodromy matrix (\ref{mon}) (with $K^{-}=1$) in the
auxiliary space:

\begin{eqnarray}
t_{q}(v)=\sum_{a}M_{aa}U_{aa}(v)=\sum_{abc}M_{ab}T_{bc}(v)T^{-1}_{ca}(-v).
\label{mtt}
\end{eqnarray}

This interpretation has been further developed also in \cite{kk}.
In fact it is known \cite{bj} that for all solution to the Yang-Baxter
equation corresponding to a non exceptional affine Lie algebra $g^{(k)}$  in
the fundamental representation, PT simmetry, unitarity and equation
(\ref{sm}) are obeyed. Also, for all cases except $D_{n}^{(2)}$, $K^{-}(v)=1$
is a solution to eqn. (\ref{sk1}) and using the automorfism (\ref{aut})
 $K^{+}(v)=M$ is a solution to eqn. (\ref{sk2}). Then for all this Lie
algebras, except $D_{n}^{(2)}$ the transfer matrix (\ref{mtt}) forms a
 commuting family,
see \cite{mn}. This transfer matrix commutes with all the generators of the
quantum algebra $U_{q}(g_{0})$ where $g_{0}$ is the maximal
finite-dimensional subalgebra of $g^{(k)}$ \cite{ksk,mn}.\\
On the other hand it can be shown in general \cite{n} that:

\begin{eqnarray}
[S,M\otimes M]=0,\nonumber
\end{eqnarray}
and,

\begin{eqnarray}
tr_{2}(M_{2}PS_{12})\propto \1,\nonumber\\
tr_{2}(M_{2}(PS)_{12}^{-1})\propto \1,\nonumber
\end{eqnarray}
giving to the transfer matrix (\ref{mtt}) the interpretation of a Markov
trace. The relation between reflection equations an Markov traces has also
been pointed out in \cite{ro1}.

\end{section}

\begin{section}{Algebraic Bethe Ansatz}
We want to diagonalize the hamiltonians (\ref{htj}). Equation (\ref{th})
shows that the eigenvalues and eigenvectors of the hamiltonians with open
boundary conditions can be obtained as derivatives of those of the transfer
matrix (\ref{tr}). Then we have to solve the eigenvalue problem:

\begin{eqnarray}
t^{\alpha\beta}\Psi=\lambda\Psi,\label{evp}
\end{eqnarray}

where $t^{\alpha\beta}(v)$ will denote the transfer matrix constructed using
family $\alpha$ of solutions for the $K^{+}$ and family $\beta$
 for $K^{-}$, $\alpha,\beta=a$ or $b$.\\
We will find for the transfer matrices in the preceding section a
generalization of the nested Bethe ansatz, only found previously for quantum
group invariant conditions \cite{fk1,suq}.
As Bethe ansatz for the transfer matrix eigenvectors,
a linear combination of $\B_a$'s acting on a  ferromagnetic ground state
 state and summed over the indices $a$ is proposed. Then one should
find the coefficients in such linear combination from the eigenvalue
condition. Surprisingly enough, these coefficients turn to obey
an eigenvector problem analogous to the original one
but with a new transfer matrix. This new transfer matrix is
built from statistical weights obtained from the original ones
deleting the first row and column. The reflection matrices give also new
reflection matrices for the reduced problem. The problem  is solved
in the sense that it reduces to a set of algebraic equations : the nested
Bethe Ansatz  equations (NBAE).\\
To follow these steps it is
necessary to make use of the commutation relations for the operators
$U_{ab}$ given in equation (\ref{sku1}) which are the same whatever
the matrix $K^{-}$ is if it obeys (\ref{sk1}) and $T$, $T^{-1}$ are
Yang-Baxter operators as defined previously. In order to simplify this
commutation relations instead of using the operators $\D_{ab}$ it is more
convenient to work with
 new operators $\d_{ab}$ such that \cite{fk1}:

\begin{eqnarray}
\D_{ab}(v)&=&\chi(v)\frac{S^{ac}_{db}(2v+\gamma)}{b(2v+\gamma)}
\d_{dc}(v)+\delta_{ab}\frac{c_{+}(2v)}{a(2v)}\A(v)\;,\label{cam}\\
\chi(v)&=&q\frac{b(2v)a(2v+\gamma)}{a(2v)b(2v+\gamma)}.\nonumber
\end{eqnarray}

Using this change and eqns. (\ref{u},\ref{sku1}) the commutation
 relations are
obtained as \cite{fk1}:

\begin{eqnarray}
\A(v)\B_{a}(v^{\prime})&=&\frac{a(v^{\prime}-v)b(v^{\prime}+v)}
{b(v^{\prime}-v)a(v^{\prime}+v)}\B_{a}(v^{\prime})\A(v)\nonumber\\
&-&\frac{c_{+}(v^{\prime}-v)b(2v^{\prime})}{b(v^{\prime}-v)a(2v^{\prime})}
\B_{a}(v)\A(v^{\prime})\nonumber\\
&-&\frac{c_{-}(v^{\prime}+v)}{a(v^{\prime}+v)}\chi(v^{\prime})
\frac{S^{bc}_{da}(2v^{\prime}+\gamma)}{b(2v^{\prime}+\gamma)}
\B_{b}(v)\d_{dc}(v^{\prime}),\label{cab}\\
\d_{bd}(v)\B_{a}(v^{\prime})&=&\frac{S^{ce}_{fd}(v-v^{\prime})}
{b(v-v^{\prime})}
\frac{S^{-1fb}_{\;\;\;\;ag}(-v-v^{\prime}-\gamma)}{\b(-v-v^{\prime}-\gamma)}
\B_{c}(v^{\prime})\d_{ge}(v)\nonumber\\
&+&\frac{1}{\chi(v)}\frac{c_{+}(v+v^{\prime})b(2v^{\prime})}
{a(v+v^{\prime})a(2v^{\prime})}\delta_{ab}\B_{b}(v)\A(v^{\prime})\nonumber\\
&-&\frac{\chi(v^{\prime})}{\chi(v)}\frac{c_{+}(v-v^{\prime})}
{b(v-v^{\prime})}
\frac{S^{bf}_{ea}(2v^{\prime}+\gamma)}{b(2v^{\prime}+\gamma)}
\B_{d}(v)\d_{ef}(^{\prime}).\label{cdb}
\end{eqnarray}
where all indexes in eqns. (\ref{cam},\ref{cab},\ref{cdb}) assume only the
values $2$ and $3$.\\
Using the change (\ref{cam}) and the definition of the transfer matrix
(\ref{tr}) it can be written as:

\begin{eqnarray}
&&t^{a\beta}=\frac{q}{\sin\xi_{+}}\left(\sin(\xi_{+}-v-\gamma)e^{iv}\A(v)+
\chi(v)K^{-}_{a(1)cd}(v)\d_{dc}(v)\right),\nonumber\\
&&t^{b\beta}=\frac{1}{\sin\xi_{+}}\left(\sin(\xi_{+}-v)e^{iv}\A(v)+
\chi(v)\sin(\xi_{+}+v+\gamma)e^{-iv}K^{-}_{b(1)cd}(v)\d_{dc}(v)\right),
\nonumber
\end{eqnarray}
with $\beta=a,b$ and:

\begin{eqnarray}
K^{-}_{a(1)}(v)&=&\left(\begin{array}{cc}
q e^{iv}\sin(\xi_{+}-v-\gamma)&0\\
0& e^{-iv}\sin(\xi_{+}+v)\\
\end{array}\right),\nonumber\\
K^{-}_{b(1)}(v)&=&\left(\begin{array}{cc}
1&0\\
0&1\\
\end{array}\right).\nonumber
\end{eqnarray}

The notation will become clear below, where we will see that these are
$K^{-}$ matrices for a reduced problem corresponding to a $spl_{q}(2,1)$
algebra.\\
It is easy to find an eigenstate
of these transfer matrices, the first level pseudovacuum, given by:

\begin{eqnarray}
\Phi=\otimes^{L}_{i=1}\left(
\begin{array}{c}
1\\0\\0
\end{array}
\right).\nonumber
\end{eqnarray}

This ferromagnetic state is an eigenvector of $\A(v)$ and $\d_{dd}(v)$. Let
us look at this point.\\
It is easy to see from the definition of $\A(v)$ that:

\begin{eqnarray}
\A(v)\Phi=a^{L}(v)\a^{L}(-v)K_{1}^{-}(v).\nonumber
\end{eqnarray}

For the case of $\D_{aa}(v)$ it is necessary to make use
of the fact that\\ $T_{a1}(v)\Phi=T^{-1}_{ab}(-v)\Phi=0$ when $a\neq
b\;\;a,b\geq 2$ and commute $T_{b1}(v)T^{-1}_{1b}(-v)$ using the
Yang-Baxter equation for the monodromy matrices $T$ and $T^{-1}$. The
result is:

\begin{eqnarray}
\D_{aa}(v)\Phi=\left(K^{-}_{a}(v)-\frac{c_{+}(2v)}{a(2v)}K^{-}_{1}(v)
\right)
b^{L}(v)\b^{L}(-v)\Phi+\frac{c_{+}(2v)}{a(2v)}\A(v)\Phi.\nonumber
\end{eqnarray}

After using the formula (\ref{cam}) we obtain:

\begin{eqnarray}
\d_{cd}(v)\Phi&=&\frac{\tau(v)}{q}\Phi K^{+}_{a(1)cd}(v),\nonumber\\
\d_{cd}(v)\Phi&=&\tau(v)\Phi \sin(\xi_{-}-v-\gamma)e^{iv}K^{+}_{b(1)cd}(v),
\nonumber
\end{eqnarray}
for the families $a$ and $b$ of $K^{-}$ matrices respectively, where:

\begin{eqnarray}
K^{+}_{a(1)}(v)&=&\left(
\begin{array}{cc}
q e^{-iv}\sin(\xi_{-}+v-\gamma)&0\\
0&-q^{2} e^{iv}\sin(\xi_{-}-v-2\gamma)\end{array}\right),\nonumber\\
K^{+}_{b(1)}(v)&=&\left(
\begin{array}{cc}
1&0\\
0&-1\end{array}\right),\nonumber
\end{eqnarray}
and:

\begin{eqnarray}
\tau(v)=\frac{b^{2}(2v+\gamma)}{q a(2v+\gamma) w(2v+\gamma)\sin\xi_{-}}
\left(\frac{c_{+}(2v)}{a(2v)}-1\right) b^{L}(v)\b^{L}(-v).\nonumber
\end{eqnarray}

The notation will become clear below, where we will see that these are
$K^{+}$ matrices for the reduced problem.\\
Also it is easily seen that
$\C_{b}(v)\Phi=0$ and that $\B_{b}(v)\Phi$ $(b=2,3)$ is not proportional
to $\Phi$ and
different from zero. Then we can use linear combinations of these
last operators to create excitations over the first level pseudovacuum in
order to look for the eigenvectors of the transfer matrices. Then we will
use for the first level Bethe ansatz the vectors:

\begin{eqnarray}
\Psi=\B_{i_{1}}(v_{1})\B_{i_{2}}(v_{2})\ldots\B_{i_{N}}(v_{N})\Phi
\Psi_{(1)}^{(i)}\;,
\nonumber
\end{eqnarray}
whith subindexes running from $2$ to $3$. The coefficients
$\Psi_{(1)}^{(i)}$ will be determined by the second-level Bethe ansatz.\\
Commuting this vector with the operators of the transfer matrices and using
the rules given by formulas (\ref{cab},\ref{cdb}) the result obtained is:

\begin{eqnarray}
t^{\alpha\beta}(v)\Psi&=&
\Gamma^{\alpha\beta}(v)\prod_{i=1}^{N}\frac{a(v_{i}-v)b(v_{i}+v)}
{b(v_{i}-v)a(v_{i}+v)}a^{L}(v)\a^{L}(-v)\Psi\nonumber\\
&+&\Delta^{\alpha\beta}
\frac{b(2v)b(2v+\gamma)}{a(2v)w(2v+\gamma)}
\left(\frac{c_{+}(2v)}{a(2v)}-1\right)
\prod_{i=1}^{N}\left(\frac{1}{b(v-v_{i})\b(-v-v_{i}-\gamma)}\right)\times
\nonumber\\
&&b^{L}(v)\b^{L}(-v)\;\;\;\B_{j_{1}}(v_{1})\B_{j_{2}}(v_{2})
\ldots\B_{j_{N}}(v_{N})\Phi
t^{\alpha\beta}_{(1)}(v+\gamma/2,\{v_{i}+\gamma/2\})^{(j)}_{(i)}
\Psi^{(i)}_{(1)}\nonumber\\
&+&u.t.\;\;\; ,\label{pi}
\end{eqnarray}

with:

\begin{eqnarray}
\Gamma^{aa}(v)&=&\Gamma^{ab}(v)=\frac{1}{\sin\xi_{+}\sin\xi_{-}}
\sin(\xi_{+}-v-\gamma)\sin(\xi_{-}+v) ,
\nonumber\\
\Gamma^{ba}(v)&=&\Gamma^{bb}(v)=\frac{1}{\sin\xi_{+}\sin\xi_{-}}
\sin(\xi_{+}-v)\sin(\xi_{-}+v),
\nonumber\\
\Delta^{aa}(v)&=&\frac{q^{-1}}{\sin\xi_{+}\sin\xi_{-}},\nonumber\\
\Delta^{ab}(v)&=&\frac{1}{\sin\xi_{+}\sin\xi_{-}}
\sin(\xi_{-}-v-\gamma) e^{iv},\nonumber\\
\Delta^{ba}(v)&=&\frac{q^{-1}}{\sin\xi_{+}\sin\xi_{-}}
\sin(\xi_{+}+v+\gamma) e^{-iv},\nonumber\\
\Delta^{bb}(v)&=&\frac{1}{\sin\xi_{+}\sin\xi_{-}}
\sin(\xi_{+}+v+\gamma)\sin(\xi_{-}-v-\gamma).\nonumber
\end{eqnarray}

The wanted terms are obtained after using the first term of the
commutation rules (\ref{cab},\ref{cdb}). The unwanted terms $u.t$ come from
the second and third terms of the commutation rules, for a detailed study of
how to deal these terms see \cite{suq}. There is no summation over the
indexes $\alpha,\beta$.
 The operator $t^{\alpha\beta}_{(1)}(v+\gamma/2,\{v_{i}+\gamma/2\})$ is the
second level transfer matrix coming from the family $\alpha$ of $K^{+}$
matrices and $\beta$ of $K^{-}$. The matrices $K^{+}_{(1)}$ and
$K_{(1)}^{-}$ turn out to be the corresponding reflection matrices for this
reduced second level problem, this explains the notation. The second level
transfer matrix is then given by:

\begin{eqnarray}
t^{\alpha\beta}_{(1)}(v+\gamma/2,\{v_{i}+\gamma/2\})=
\sum_{c=2}^{3}K^{+}_{\beta(1)cc}(v)
U_{\alpha(1)cc}(v+\gamma/2,\{v_{i}+\gamma/2\}),\nonumber
\end{eqnarray}
where the second level $U_{\alpha(1)}(v,\{v_{i}\})$ operator is formed
from the weights given in the first term in (\ref{cdb}) and the
$K^{-}_{\alpha(1)}$ matrix.\\
We are led from equation (\ref{pi}) to a new eigenvalue problem for a
reduced transfer matrix with only two states per link corresponding to the
superalgebra $sl_{q}(1,1)$:

\begin{eqnarray}
t^{\alpha\beta}_{(1)}(v+\gamma/2,\{v_{i}+\gamma/2\})\Psi_{(1)}=
\lambda^{\alpha\beta}_{(1)}(v)\Psi_{(1)}.\label {zz}
\end{eqnarray}

It can be seen that the unwanted terms in (\ref{pi}) cancel if:

\begin{eqnarray}
& &\frac{\Gamma^{\alpha\beta}}{\Delta^{\alpha\beta}}\prod_{i\neq k}^{N}
\frac{a(v_{i}-v_{k})b(v_{i}+v_{k})}{b(v_{i}-v_{k})a(v_{i}+v_{k})}
a^{L}(v_{k})\a^{L}(-v_{k})+\nonumber\\
&&\frac{\lambda^{\alpha\beta}_{(1)}(v_{k})}
{\sin\gamma\;a(2v_{k}+\gamma)}\left(\frac{c_{+}(2v_{k})}{a(2v_{k})}-1\right)
\nonumber\\
&&\prod_{i\neq k}^{N}\left(\frac{1}{b(v_{k}-v_{i})\b(-v_{k}-v_{i}-\gamma)}
\right) b^{L}(v_{k})\b^{L}(-v_{k})=0,
\;\;\;k=1,...,N\label{hw1}
\end{eqnarray}
notice that the factor depending on
$\alpha,\beta$ is equal to $1$ in the quantum group invariant
case.\\
To solve the reduced problem we can follow parallel
steps from those of the previous level. First define the operators
$\A_{(1)}=U_{(1)22}$,  $\B_{(1)}=U_{(1)23}$, $\C_{(1)}=U_{(1)32}$ and
$\D_{(1)}=U_{(1)33}$. The second level Bethe ansatz for $\Psi_{1}$ is
given by:

\begin{eqnarray}
\Psi_{(1)}=\B_{(1)}\left(\nu_{1}+\gamma/2,\{v_{i}+\gamma/2\}\right)
\B_{(1)}\left(\nu_{2}+\gamma/2,\{v_{i}+\gamma/2\}\right)\ldots\nonumber\\
\B_{(1)}\left(\nu_{M}+\gamma/2,\{v_{i}+\gamma/2\}\right)\Phi_{(1)},\nonumber
\end{eqnarray}
where:

\begin{eqnarray}
\Phi_{(1)}=\otimes^{N}_{i=1}\left(
\begin{array}{c}
1\\0
\end{array}
\right),\nonumber
\end{eqnarray}
is the second level pseudo vacuum. This is anihilated by the $\C_{(1)}$
operator. As in the first level it is convenient to make the change:

\begin{eqnarray}
\d_{(1)}(v,\{v_{i}\})=\D_{(1)}(v,\{v_{i}\})-\frac{c_{-}(2v)}{a(2v)}
\A_{(1)}(v,\{v_{i}\}).\label{cam1}
\end{eqnarray}

The action of $\A_{(1)}$ and $\d_{(1)}$ on $\Phi_{(1)}$ is given by:

\begin{eqnarray}
\A_{(1)}(v+\gamma/2,\{v_{i}\gamma/2\})\Phi_{(1)}&=&
K^{-}_{(1)11}(v)\prod_{i=1}^{N}a(v-v_{i})\a(-v-v_{i}-\gamma)\Phi_{(1)}
\label{ap}\\
\d_{(1)}(v+\gamma/2,\{v_{i}\gamma/2\})\Phi_{(1)}&=&
\mu_{\alpha}(v)q\frac{b(2v+\gamma)}{a(2v+\gamma)}
\prod_{i=1}^{N}b(v-v_{i})\b(-v-v_{i}-\gamma)\Phi_{(1)}\label{pp}\\
\mu_{a}(v)&=&q^{-1}e^{-iv}\sin(\xi_{+}+v+\gamma)\nonumber\\
\mu_{b}(v)&=&1.\nonumber
\end{eqnarray}

The reduced transfer matrices can be written after the change (\ref{cam1}):

\begin{eqnarray}
t_{(1)}&=&\theta_{\alpha}(v)q\frac{b(2v+\gamma)}{a(2v+\gamma)}\A_{(1)}-
\rho_{\alpha}(v)\d_{(1)}\label{uu}\\
\theta_{a}(v)&=&e^{-iv}\sin(\xi_{-}+v),\;\;\;\theta_{b}(v)=1\nonumber\\
\rho_{a}(v)&=&q^{2}e^{iv}\sin(\xi_{-}-v-2\gamma),\;\;\;
\rho_{b}(v)=1.\nonumber
\end{eqnarray}

The commutation relations for the operators $\A_{(1)}$, $\d_{(1)}$
and $\B_{(1)}$ follow from (\ref{sku1}), see \cite{fk1}. Using these and
(\ref{uu},\ref{ap},\ref{pp}) a reasoning similar to the first level Bethe
ansatz gives the final result for the eigenvalue problem (\ref{evp}):

\begin{equation}
\lambda(v)=\lambda_{\A}(v)+\lambda_{\D_{I}}(v)+\lambda_{\D_{II}}(v),
\label{solt}
\end{equation}
with:

\begin{eqnarray}
\lambda_{\A}(v)&=&\lambda_{(\alpha\beta)\A}(v)
\prod_{i=1}^{N}\frac{a(v_{i}-v)b(v_{i}+v)}
{b(v_{i}-v)a(v_{i}+v)}a^{L}(v)\a^{L}(-v),\\
\nonumber\\
\lambda_{(aa)\A}(v)&=&\lambda_{(ab)\A}(v)=
\frac{q\sin(\xi_{+}-v-\gamma)\sin(\xi_{-}+v)}{\sin\xi_{+}\sin\xi_{-}},
\nonumber\\
\lambda_{(ba)\A}(v)&=&\lambda_{(bb)\A}(v)=
\frac{q\sin(\xi_{+}-v)\sin(\xi_{-}+v)}{\sin\xi_{+}\sin\xi_{-}},\nonumber\\
\nonumber\\
\lambda_{\D_{I}}(v)&=&\lambda_{(\alpha\beta)\D_{I}}(v)
\frac{b(2v)b(2v+\gamma)}{a(2v)w(2v+\gamma)}
\left(\frac{c_{+}(2v)}{a(2v)}-1\right)
\left(1-\frac{c_{-}(2v+\gamma)}{a(2v+\gamma)}\right)
b^{L}(v)\b^{L}(-v)\nonumber\\
&&\prod_{i=1}^{N}\frac{a(v-v_{i})\a(-v-v_{i}-\gamma)}
{b(v-v_{i})\b(-v-v_{i}-\gamma)}
\prod_{j=1}^{M}\frac{a(\nu_{j}-v)b(\nu_{j}+v+\gamma)}
{b(\nu_{j}-v)a(\nu_{j}+v+\gamma)}\\
\nonumber\\
\lambda_{(aa)\D_{I}}(v)&=&\frac{q\sin(\xi_{+}-v-\gamma)\sin(\xi_{-}+v)}
{\sin\xi_{+}\sin\xi_{-}}\nonumber\\
\lambda_{(ab)\D_{I}}(v)&=&\frac{q^{2}\sin(\xi_{-}-v-\gamma)
\sin(\xi_{+}-v-\gamma)e^{2iv}}{\sin\xi_{+}\sin\xi_{-}}\nonumber\\
\lambda_{(ba)\D_{I}}(v)&=&\frac{q^{-1}\sin(\xi_{+}+v+\gamma)
\sin(\xi_{-}+v)e^{-2iv}}{\sin\xi_{+}\sin\xi_{-}}\nonumber\\
\lambda_{(bb)\D_{I}}(v)&=&\frac{\sin(\xi_{+}+v+\gamma)
\sin(\xi_{-}-v-\gamma)}{\sin\xi_{+}\sin\xi_{-}}\nonumber\\
\nonumber\\
\lambda_{\D_{II}}(v)&=&-\lambda_{(\alpha\beta)\D_{II}}(v)
\frac{b(2v)b(2v+\gamma)}{a(2v)w(2v+\gamma)}
\left(\frac{c_{+}(2v)}{a(2v)}-1\right)
\left(1-\frac{c_{-}(2v+\gamma)}{a(2v+\gamma)}\right)
\nonumber\\
&&b^{L}(v)\b^{L}(-v)\prod_{j=1}^{M}\frac{a(\nu_{j}-v)b(\nu_{j}+v+\gamma)}
{b(\nu_{j}-v)a(\nu_{j}+v+\gamma)}\\
\nonumber\\
\lambda_{(aa)\D_{II}}(v)&=&\lambda_{(ba)\D_{II}}(v)=
\frac{q\sin(\xi_{+}+v+\gamma)\sin(\xi_{-}-v-2\gamma)}
{\sin\xi_{+}\sin\xi_{-}}\nonumber\\
\lambda_{(bb)\D_{II}}(v)&=&\lambda_{(ab)\D_{II}}(v)=
\frac{\sin(\xi_{-}-v-\gamma)
\sin(\xi_{+}+v+\gamma)}{\sin\xi_{+}\sin\xi_{-}}\nonumber
\end{eqnarray}

In the previous equations it is easily seen that for the case
$(\alpha,\beta)=(b,b)$ the eigenvalue is real. For the case
$(\alpha,\beta)=(a,a)$ there is an overall factor $q$ which can be eliminated
by the redefinition $K^{+}_{a}\rightarrow q^{-1}K^{+}_{a}$, making the
eigenvalue real, (we have maintained the present definition in order to make
more clear the quantum group invariant limit). For the ``mixed" cases
$(a,b),(b,a)$, we have imaginary terms which are not possible to eliminate
by redefinitions of the reflection matrices , or by gauge transformations of
the $S$ matrix. As explained in sect. 2 the eigenvalues are real for all
the cases in the hiperbolic and rational regimes.\\
The limit
$\Xi_{\pm}\rightarrow\infty$ leads to the formulas obtained in \cite{fk1}
for the quantum group invariat case.\\
The nested Bethe ansatz equations
obtained from the cancelation of unwanted terms are:

\begin{eqnarray}
&&\eta_{\alpha\beta}(v_{k})\left(\frac{a(v_{k})\a(-v_{k})}{b(v_{k}\b(-v_{k})}
\right)^{L}
\prod_{i\neq k}^{N}\frac{a(v_{i}-v_{k})b(v_{i}+v_{k})\b(v_{i}+v_{k}+\gamma)}
{a(v_{k}-v_{i})a(v_{i}+v_{k})\a(-v_{i}-v_{k}-\gamma)}\nonumber\\
&&\prod_{j=1}^{M}\frac{a(\nu_{j}+v_{k}+\gamma)b(\nu_{j}-v_{k})}
{b(\nu_{j}+v_{k}+\gamma)a(\nu_{j}-v_{k})}=1,\;\; k=1,...,N\;,
\label {baeq1}\\
&&\zeta_{\alpha\beta}(v_{k})
\prod_{i=1}^{N}\frac{a(\nu_{l}-v_{i})\a(-\nu_{l}-v_{i}-\gamma)}
{b(\nu_{l}-v_{i})\b(-\nu_{l}-v_{i}-\gamma)}=1,\;\; l=1,...M\label{baeq}\\
&&\begin{array}{cc}
\eta_{aa}(v_{k})=1&
\zeta_{aa}(v_{k})=\frac{\sin(\xi_{-}+\nu_{l})
\sin(\xi_{+}-\nu_{l}-\gamma)}{\sin(\xi_{-}-\nu_{l}-2\gamma)
\sin(\xi_{+}+\nu_{l}+\gamma)}\\
\eta_{ab}(v_{k})=\frac{\sin(\xi_{-}+v_{k})e^{-2iv_{k}}}
{q\sin(\xi_{-}-v_{k}-\gamma)}& \zeta_{ab}(v_{k})=
\frac{q^{2}\sin(\xi_{+}-\nu_{l}-\gamma)e^{2i\nu_{l}}}
{\sin(\xi_{+}+\nu_{l}+\gamma)}\\
\eta_{ba}(v_{k})=\frac{q\sin(\xi_{+}-v_{k})e^{2iv_{k}}}
{\sin(\xi_{+}+v_{k}+\gamma)}&\zeta_{ba}(v_{k})=
\frac{\sin(\xi_{-}+\nu_{l})e^{-2i\nu_{l}}}
{q^{2}\sin(\xi_{-}-\nu_{l}-2\gamma)}\\
\eta_{bb}(v_{k})=\frac{\sin(\xi_{+}-v_{k})\sin(\xi_{-}+v_{k})}
{\sin(\xi_{+}+v_{k}+\gamma)\sin(\xi_{-}-v_{k}-\gamma)}&
\zeta_{ba}(v_{k})=1\end{array}\nonumber
\end{eqnarray}

Then the solution to the eigenvalue problem (\ref{evp}) is given by
equation (\ref{solt}), with $v_{i},\;\nu_{l}$ given by equations
(\ref{baeq1},\ref{baeq}).\\
This  equations are real for
$(\alpha,\beta)=(a,a),(b,b)$ and imaginary for the "mixed" cases. This
new fact is present only in the case of independent open boundary conditions
for $S$-matrices of algebras with rank larger than one. This is the first
model solved for these conditions and this property is also expected
in other systems \cite{suq,ks}. This terms may deserve further study in the
hyperbolic regime, but for the trigonometric regime do not seem to have
physical applications. In the quantum group invariant limit all these terms
dissapear. Imaginary eigenvalues appear also in the periodic case for this
$S$ matrix as noticed in \cite{fk1}.

\end{section}

\begin{section}{Highest weight property for the $spl_{q}(2,1)$ invariant
t-J model}

In this section the highest weight property for the Bethe
states of the quantum group invariant case is proved. This property has been
 also shown to hold for the Bethe states of the $SU_{q}(2)$ invariant XXZ
chain \cite{qba1,mn2}, but it is shown for the first time for a quantum group
invariant chain with more than two states per link.\\
In reference \cite{fk1} a representation in the lattice  of the generators
of $spl_{q}(2,1)$ was obtained at certain limits of the spectral parameter
$v$. The important ones for what follows are:

\begin{eqnarray}
\A(x\rightarrow\infty)&\sim& q^{-L}q^{2W_{1}},\nonumber\\
\D_{33}(x\rightarrow 0)&\sim&q^{L}q^{2W_{3}},\nonumber\\
\C_{2}(x\rightarrow\infty)&\sim&\alpha_{-}q^{-L/2}q^{-W_{3}/2}F_{1}q^{W_{1}}
,\nonumber\\
\D_{32}(x\rightarrow 0)&\sim&-\alpha_{+}q^{L/2}\tilde{\sigma}
q^{W_{1}/2+2W_{3}}F_{2},\label{lim}\\
\alpha_{\pm}&=&q^{\pm 1/2}(q-q^{-1}),\nonumber\\
\tilde{\sigma}&=&\sigma\otimes\sigma\otimes\ldots\otimes\sigma,\;\;\;\;
\sigma=diag(1,1,-1).\nonumber
\end{eqnarray}

In the previous formulas $F_{1}$, $F_{2}$ are generators of the quantum
algebra, the rest of generators are $E_{1}$, $E_{2}$ , $H_{1}=W_{1}-W_{2}$
and  $H_{2}=W_{2}+W_{3}$. These obey:

\begin{eqnarray}
q^{H_{1}}q^{H_{2}}&=&q^{H_{2}}q^{H_{1}},\nonumber\\
q^{H_{i}}F_{j}q^{-H_{i}}&=&q^{a_{ij}}F_{j},\nonumber\\
q^{H_{i}}E_{j}q^{-H_{i}}&=&q^{a_{ij}}E_{j},\nonumber\\
{[}F_{1},E_{1}]&=&\frac{q^{H_{1}}-q^{-H_{1}}}{q-q^{-1}},
\;\;[F_{1},E_{2}]=0,\nonumber\\
{[} F_{2},E_{2}]_{+}&=&\frac{q^{H_{2}}-q^{-H_{2}}}{q-q^{-1}},
\;\;[ F_{2},E_{1}]=0,\nonumber\\
E_{2}^{2}&=&F^{2}_{2}=0,\nonumber
\end{eqnarray}
plus $q$-Serre relations. Here $a_{ij}$ are the elements of the graded
Cartan matrix given by $a_{11}=2,a_{12}=a_{21}=-1,a_{22}=0$. These are the
commutation relations defining the quantum group $spl_{q}(2,1)$. These
generators can be shown to commute with the quantum group invariant transfer
matrix $t(x)$ using the method in  references \cite{ksk,mn} or by  direct
calculation \cite{fk1}, i.e:

\begin{eqnarray}
[t(x),q^{H_{i}}]&=&0, \nonumber\\
{[} t(x),F_{i}]&=&0,\nonumber\\
{[}t(x),E_{i}]&=&0,\;\; i=1,2.\label{mas}
\end{eqnarray}

To prove the highest weight property for the Bethe states $\Psi$ it is
neccesary to show that $F_{1}\Psi=F_{1}\Psi=0$. For that we need to
know the commutation relations between the operators $\B_{d}$ and the
generators $F_{i}$. These are obtained using equations
(\ref{lim},\ref{sku1})
 and
making the neccesary limits. The result for $F_{1}$ is:

\begin{eqnarray}
F_{1}\B_{d}(v)&=&q^{(2-d/2)}\B_{d}(v)F_{1}+
q^{\frac{(1-L)}{2}}q^{-W_{3}/2+W_{1}}\times
\nonumber\\
&&\left(\delta_{d2}\left(1-\frac{c_{+}(2v)}{a(2v)}\right)\A(v)-\chi(v)
\frac{S_{bd}^{2c}(2v+\gamma)}{b(2v+\gamma}\d_{bc}(v)\right),\nonumber
\end{eqnarray}
where we have also used (\ref{cam}) and the commutation relations:

\begin{eqnarray}
[q^{W_{1}},\A(x)]=0,\;[W_{3},\B_{d}(v)]=\delta_{d3}\B_{d}(v),
\;[W_{1},\B_{d}(v)]=-\B_{d}(v).\nonumber
\end{eqnarray}

Using the previous results, the commutation relations (\ref{cab},\ref{cdb})
and
the fact that $F_{1}\Phi=0$ we find:

\begin{eqnarray}
F_{1}\Psi&=&\sum_{i=1}^{N}q^{p}\delta_{i_{k}2}\frac{b(2v_{k})}{a(2v_{k})}
\times\nonumber\\
&\{&\prod_{i\neq k}^{N}
\frac{a(v_{i}-v_{k})b(v_{i}+v_{k})}{b(v_{i}-v_{k})a(v_{i}+v_{k})}
a^{L}(v_{k})\a^{L}(-v_{k})\nonumber\\
&+&\frac{\lambda_{(1)}(v_{k})}
{\sin\gamma\;a(2v_{k}+\gamma)}\left(\frac{c_{+}(2v_{k})}{a(2v_{k})}-1\right)
\prod_{i\neq k}^{N}\left(\frac{1}{b(v_{k}-v_{i})\b(-v_{k}-v_{i}-\gamma)}
\right)
\times\nonumber\\
&&b^{L}(v_{k})\b^{L}(-v_{k})\}\;
\B_{i_{k+1}}(v_{k})\ldots\B_{i_{k-1}}(v_{k-1})
\Phi M^{(i)}_{(j)}\Psi^{(j)}_{(1)},\nonumber
\end{eqnarray}
where $p$ is an operator irrelevant for what follows, and we have made use
of equation (\ref{zz}). The matrix  $M^{(i)}_{(j)}$ takes count of the
reordering of the $\B$ operators see \cite{hr,suq}. We see looking to the
equation (\ref{hw1}) that this last expression shows that $F_{1}\Psi=0$.\\
For the operator $F_{2}$ we use the relation (\ref{cam})to find that:

\begin{eqnarray}
\d_{32}(x\rightarrow 0)=-q^{\frac{L-1}{2}}(q-q^{-1})\tilde{\sigma}
q^{W_{1}/2+2W_{3}}F_{2}.\nonumber
\end{eqnarray}

Using the commutation relations (\ref{cdb}) in the limit
$x\rightarrow 0$ we see that the first summand is of order $x^{-2}$
with respect to the third and fourth and is the only one which survives in
this limit. Using the obtained commutation relation N times:

\begin{eqnarray}
F_{2}\Psi=c \B_{j_{1}}(v_{1})\ldots\B_{j_{N}}(v_{N})\Phi
F^{\;\;\;\{j\}}_{(1)2\{i\}}\Psi^{\{i\}}_{(1)},\nonumber
\end{eqnarray}
where $F_{(1)2}$ is the $x\rightarrow 0$ limit of $U_{(1)23}$ and $c$
is an unimportant factor different from zero. We use again the relation
(\ref{sku1}) for the reduced problem and take the corresponding limits to
find:

\begin{eqnarray}
F_{(1)2}\B_{(1)}(v)=q^{2}\B_{(1)}(v)F_{(1)2}+(q-q^{-1})\D_{(1)}(x=0)
\left(\D_{(1)}(v)-\A_{(1)}(v)\right),\nonumber
\end{eqnarray}
where we have also used $[\D_{(1)}(x=0),\A_{(1)}(v)]=0$. Using the
previous commutation relation, the change (\ref{cam1}) and the fact that
$F_{(1)2}\Phi_{(1)}=0$ we find:

\begin{eqnarray}
F_{(1)2}\Psi_{(1)}&=&\tilde{c} \sum_{l=1}^{M}\prod_{j\neq l}^{M}
\frac{a(\nu_{j}-\nu_{l})b(\nu_{j}+\nu_{l}+\gamma)}
{b(\nu_{j}-\nu_{l})a(\nu_{j}+\nu_{l}+\gamma)}
\left(1-\frac{c_{-}(2\nu_{l}+\gamma)}{a(2\nu_{l}+\gamma)}\right)\times
\nonumber\\
&&\left\{\prod_{i=1}^{N} b(\nu_{l}-v_{i})\b(-\nu_{l}-v_{i}-\gamma)-
\prod_{i=1}^{N}
a(\nu_{l}-v_{i})\a(-\nu_{l}-v_{i}-\gamma)\right\}\times
\nonumber\\
&&\B_{(1)}(v_{l+1})\ldots\B_{(1)}(v_{l-1})\Phi_{(1)},\nonumber
\end{eqnarray}
where $\tilde{c}$ is an  operator that does not affect the following
argument. It is clear, after looking to formula (\ref{baeq}) in the quantum
group invariant limit, that $F_{2}\Psi=0$. This finish the proof of the
highest weight property for the Bethe eigenvectors of the $spl_{q}(2,1)$
invariant t-J chain. As the kernel of $F_{1}$, $F_{2}$ is stable under
variations of $\gamma$, the highest weight property holds even when the
$\gamma/\pi$ is rational; this can be important for the study of the
associated RSOS models \cite{qba2}.\\
{}From equations (\ref{mas}) we can see that the eigenvectors are
classified by multiplets corresponding to irreducible representations of
the quantum superalgebra $spl_{q}(2,1)$. The eigenvalues are
degenerate, since the vectors
$E_{i}\Psi,E_{i}^{2}\Psi,...,E_{i}^{J}\Psi,\;\;\; i=1,2$, with $J\neq 0$ in
general,  are all eigenvectors of $t(v)$ with the same eigenvalue. All
these vectors are included in the same multiplet whose highest weight
vector is given by a Bethe ansatz state. These vectors are not comming from
the Bethe ansatz but all are directly obtained from it by applying the,
lowering, $E_{i}$ operators. To deal with the problem of completeness it
would be neccesary to  to find
a general expresion for the degeneracy of the eigenvalues and study the
Bethe ansatz equations, and then to make an analysis paralell to that of the
$spl(2,1)$ invariant t-J model \cite{fk2}.

\end{section}

\begin{section}{Conclusions}
We have presented open boundary conditions compatible with integrability
for the one dimensional supersymmetric t-J model. It turns
out that there are four families of boundary conditions commuting with
$n$ and $S^{z}$ depending each one on two arbitrary parameters. A
connection betwen the Markov trace method for finding open boundary
conditions and reflection equations have been found, showing in this way its
generality. We  have also found the relation between the one dimensional t-J
model  with open boundary conditions and the open transfer matrix of the
graded 15 vertex model. The models proposed have been diagonalized using a
generalization of the quantum group invariant nested Bethe ansatz for this
kind of boundary conditions. The $spl_{q}(2,1)$ invariant chain is obtained
as a special limit. It turns out that for some ''mixed" cases the eigenvalues
in the trigonometric regime are imaginary. This phenomenon is new and
characteristic of open boundary conditions for systems with more than two
states per link.\\
The highest weight property for the Bethe states of the
quantum group invariant t-J model have been shown. Using the quantum group
invariance of the chain we have found that the eigenvectors are classified
by multiplets. These multiplets of highest weight given by the Bethe
ansatz vectors are generated by the lowering $E_{i}$ operators applied to
the corresponding Bethe sate. We hope that this combination of Bethe
ansatz and the quantum group properties of the model will give a
complete set of eigenvectors as is the case for the $spl(2,1)$ periodic
chain \cite{fk2}.\\
It would be interesting to make a parallel study for
models with open boundary conditions associated to the algebras $A_{n-1}$.
This question has been recently addressed in \cite{hwp,last}. The problem of
completeness remains open, and further investigation on the quantum group
properties  of the model and the Behte ansatz equations could bring some
advance. The proof of the highest weight property  opens the study of the
associated RSOS models.

\end{section}

\vspace{3cm}

I would like to thank H.J. de Vega, R. Cuerno and F. Guil for helpful
discussions and comments, and P.P. Kulish for focushing my attention to the
subject.

\end{document}